\documentclass[a4paper,12pt]{article}
\usepackage[cp1251]{inputenc}
\usepackage[english]{babel}
\usepackage[dvips]{graphicx}
\graphicspath{{.}}

\begin{document}
\mathsurround=2pt \sloppy
\title{\bf  Phenomenological phase diagram of superfluid $^3$He in a stretched aerogel}
\author{I. A. Fomin
\vspace{.5cm}\\
{\it P. L. Kapitza Institute for Physical Problems}\\ {\it Russian Academy of Science},\\{\it
Kosygina 2,
 119334 Moscow, Russia}}
\maketitle
\begin{abstract}
Highly anisotropic "nematically ordered" aerogel induces global uniaxial anisotropy in the superfluid $^3$He. The anisotropy lowers symmetry of $^3$He in aerogel from spherical to axial. As a result, instead of one transition temperature in a state with orbital moment $l$=1 there are two, corresponding to projections $l_z$=0 and $l_z=\pm$1. This splitting has a pronounced effect on the phase diagram of superfluid $^3$He and on structures of appearing phases. Possible phase diagrams, obtained phenomenologically on a basis of Landau expansion of thermodynamic potential in a vicinity of the transition temperature are presented here. The order parameters, corresponding to each phase and their temperature dependences are found.
\end{abstract}

\section{Introduction}
At a triplet Cooper pairing transition temperature $T_c$ is degenerate with respect to three projections of spin. In the superfluid $^3$He, where Cooper pairs are formed in a state with the orbital angular momentum $l=1$ there is additional degeneracy with respect to 3 projections of orbital angular momentum. A proper superposition of all components is represented by the order parameter, which is a 3$\times$3 matrix of complex amplitudes $A_{\mu j}$.   The spin projections are labeled here by index $\mu$  and orbital -- by $j$. The concrete  form of the order parameter is determined by minimization of the corresponding thermodynamic potential with respect to $A_{\mu j}$. In the case of superfluid $^3$He, depending on pressure, the stable minima correspond to order parameters describing the Anderson Brinkman Morel (ABM) or Ballian Werthammer (BW) phases \cite{VW}. In both cases a form of the order parameter does not change with the temperature, only the overall amplitude $\Delta$ grows at cooling. This is eventually a manifestation of the mentioned degeneracy.

Lowering of spherical symmetry of liquid $^3$He by external fields or oriented impurities can split transition to the superfluid state and to separate partly the components which at cooling evolve together into the corresponding order parameter.
E.g. the degeneracy of $T_c$ over spin projections is lifted by magnetic field $H_{\mu}$. Its principal effect is described by the Zeeman term in the free energy: $\Phi_H\sim H_{\mu}H_{\nu}A_{\mu j}A_{\nu j}^*$, which has to be added to the  expansion of free energy in powers of $A_{\mu j}$. As a result the transition temperature  T$_c$ is split in two, so that the temperature of transition for  $s_z=\pm 1$ is higher than that for $s_z=0$ and in magnetic field ABM-phase, which does not include $s_z=0$ component is formed first.

Similarly, degeneracy of $T_c$ over the orbital projections is lifted by a global orbital anisotropy. Such anisotropy can be induced by a deformed aerogel immersed in the superfluid $^3$He \cite{kunim}. Aoyama and Ikeda \cite{AI} considered theoretically effect of a uniaxial global anisotropy on the phase diagram of superfluid $^3$He. Their argument was based on a model, in which global anisotropy is induced by the averaged effect of anisotropic scattering of quasi-particles by oriented impurities. They predicted, in particular, that a uniaxial stretch of aerogel just below the transition temperature would stabilize the polar phase, which on cooling to lower temperatures undergoes a continuous transition to the distorted ABM-phase and eventually BW phase is formed via  the  first order transition.
These predictions were put to a test in experiments with the "nematically ordered" aerogel \cite{askh}, which can be considered as infinitely stretched. Experimentally found phase diagram confirms the predicted sequence of the phase transitions, but other, even qualitative features of the two phase diagrams are different.

In a present paper possible phase diagrams of superfluid $^3$He in a stretched aerogel are considered phenomenologically. It is shown, that depending on values of phenomenological parameters, characterizing this system different routes of development of the order parameter at cooling from the transition temperature are possible.  Orbital  anisotropy is formally described by additional term in the thermodynamic potential $\Phi_{\kappa}\sim \kappa_{jl}A_{\mu j}A_{\mu l}^*$, where $\kappa_{jl}$ is a real symmetric tensor, which can be defined as a traceless. It is assumed to be uniform (i.e. does not depend on coordinate). Random local anisotropy is neglected. This approximation is well justified in a present discussion, when only structures of order parameters of possible phases are considered. On the other hand, random anisotropy can strongly effect orientation of order parameters of the distorted ABM and of the axi-planar phases, giving rise to a randomly non-uniform Larkin-Imry-Ma (LIM) state. In the case of a  stretched aerogel it is a two dimensional LIM state, as discussed in Refs.\cite{dm2,vol1}. At comparison with experiment, in particular with NMR data a corresponding averaging over orientations of the order parameter has to be made.  With account of the global anisotropy the standard expansion of thermodynamic potential in powers of $A_{\mu j}$ is:
$$
 \Phi_s=\Phi_n+N_{eff}[(\tau\delta_{jl}+\kappa_{jl}) A_{\mu j}A_{\mu l}^*+
\frac{1}{2}(\beta_1A_{\mu j}A_{\mu j}A_{\nu l}^*A_{\nu l}^*+\beta_2A_{\mu j}A_{\mu j}^*A_{\nu l}A_{\nu l}^*+
$$
$$\beta_3A_{\mu j}A_{\nu j}A_{\mu l}^*A_{\nu l}^*+\beta_4A_{\mu j}A_{\nu j}^*A_{\nu l}A_{\mu l}^*+
\beta_5A_{\mu j}A_{\nu j}^*A_{\mu l}A_{\nu l}^*)]       \eqno(1)
$$
Here $\tau=(T-T_c)/T_c$ is the dimensionless temperature,  $T_c$ is the transition temperature, defined so that it includes all global isotropic shifts from that of bulk $^3$He. The overall coefficient $N_{eff}$ has dimensionality of a density of states. Phenomenological coefficients  $\beta_1,...\beta_5$  depend on pressure and properties of aerogel.  When anisotropy is uniaxial, in proper axes $\kappa_{xx}=\kappa_{yy}=\kappa$, $\kappa_{zz}=-2\kappa$.
In a contrast to magnetic field, which always favors $s_z=\pm 1$ projections, a uniaxial deformation of aerogel, depending on the sign of $\kappa$ favors either $l_z=\pm 1$ or $l_z=0$  projection.
For a compressed aerogel $\kappa>0$ and the states with $l_z=\pm 1$ have higher transition temperature, while for a stretched aerogel $\kappa<0$ a state with  $l_z=0$ is favored.

Stabilization of the polar phase by a stretched aerogel within this approach follows immediately from the explicit form of the
second order terms  in the expression for the thermodynamic potential (1):  $(\tau+2\kappa)A_{\mu z}A_{\mu z}^*+(\tau-\kappa)(A_{\mu x}A_{\mu x}^*+A_{\mu y}A_{\mu y}^*)$. For negative $\kappa$ the highest transition temperature is $\tau=-2\kappa$. For realistic values of coefficients $\beta$, in particular if $\beta_{15}<0$ (here and in what follows conventional shorthand notation for sums of coefficients $\beta$ are used, e.g. $\beta_1+\beta_5=\beta_{15}$ etc.) below $\tau=-2\kappa$ the superfluid polar phase is favored \cite{FS}.  Its order parameter can be written as $A_{\mu j}^{0}=\Delta_0\exp(i\varphi)d_{\mu}m_j$, where  $d_{\mu}$ is a real spin vector and $m_j$ is a unit vector in $z$-direction.

The polar phase is stable within the interval of temperatures $\tau\sim\kappa$. On further cooling the suppressed projections of angular momentum $l_z=\pm 1$ come into effect, they change a symmetry of the order parameter and further phase transitions can take place. While stabilization of the polar phase practically depends only on a sign of $\kappa$ the interval of its stability and sequence of further transitions depend also on the values of the coefficients  $\beta_1,...\beta_5$. To avoid discussion of non-realistic situations we have to restrict region of admitted values of $\beta$-s. Within the BCS theory  their values are proportional to one combination of parameters $\beta_0=\frac{7\zeta(3)}{8\pi^2T_c^2}$: $\beta_1,...\beta_5$=$\beta_0$(-1/2,1,1,1,-1). This set of values of $\beta$-s is referred as the weak coupling limit \cite{VW}. In the definition of $\beta_0$  $N(0)$ is the density of states and $\zeta(3)$ - Riemann zeta-function. The observed thermodynamic properties of bulk superfluid $^3$He in a vicinity of $T_c$ can be fitted by the $\beta_1,...\beta_5$, which deviate from their weak coupling values for 10-20\% \cite{halp2}. The deviations are smaller at low pressures. For $^3$He in aerogel situation is less certain.
Impurities give rise to corrections to the $\beta$-coefficients of the order of $\xi_0/\lambda$, where $\xi_0$ is the correlation length of superfluid $^3$He and  $\lambda$ the mean free path. This ratio is  of the order of $1/10$.
 In what follows we assume that deviations of $\beta$-s for superfluid $^3$He in nematically ordered aerogel from their weak coupling values are also of the order of $1/10$ at least at low pressures.

There is another reason for restricting the present discussion to a region of low pressures (say - below 10 bar).
The diameters of strands in nematically ordered aerogel, estimated as d$\sim$10 nm \cite{askh} are bigger than in silica aerogels and can be comparable with the correlation length of superfluid $^3$He, which at pressures above 20 bar is about 20 nm.
When $d\sim\xi_0$ perturbation of the order parameter in a vicinity of a strand is of the order of unity. Well below T$_c$ the condensate varies on a distance $\sim\xi_0$, which is smaller than the average distance between the strands $\xi_a\simeq 200$ nm. In this situation condensate is essentially nonuniform and the average order parameter does not properly characterize the state of $^3$He. Uniform approximation works better at low pressures and in a vicinity of the T$_c$ in a region where the Ginzburg and Landau (GL) coherence length $\xi(T)$ exceeds not only diameter of a strand but also $\xi_a$. In this - GL region $d\ll\xi_a\ll\xi(T)$ the average order parameter $A_{\mu j}$ is a suitable characteristic of a state of superfluid $^3$He.

Preliminary results of phenomenological analysis of the phase diagram of superfluid $^3$He in nematically ordered aerogel were
published before \cite{FS}. A principal suggestion of this paper was to consider the extra line (ESP2) in the experimentally found phase diagram as an evidence of possible stability (or meta-stability) of the axi-planar phase. Further experiments and their  analysis \cite{ESP2} have shown that this suggestion is not correct. Nevertheless there remains question of possible stability of the axi-planar phase in anisotropic environment. In Ref. \cite{FS} it was shown that the axi-planar phase is a possible minimum of thermodynamic potential at the weak coupling values of phenomenological coefficients $\beta$.  The weak coupling limit corresponds to a singular point in the space of parameters $\beta$ and solution presented in Ref.\cite{FS} is only one of many possibilities. In what follows a question of stability of the axi-planar phase is discussed with account of the mentioned singularity  and conditions, determining possibility  of its existence are found.
\section{Further phases}
To find further possible phase transitions we represent the order parameter as $A_{\mu j}=A_{\mu j}^{0}+a_{\mu j}$, where
$A_{\mu j}^{0}=\Delta_0\exp(i\varphi)d_{\mu}m_j$ is the order parameter of the polar phase and $a_{\mu j}$ -- a small increment, and expand the change of the thermodynamic potential
$\bar{\Phi}\equiv (\Phi_s-\Phi_n)/N_{eff}$ in powers of $a_{\mu j}$, separating terms of different order:
$$
\bar{\Phi}=\bar{\Phi}_0+\bar{\Phi}_2+\bar{\Phi}_4.                                                     \eqno(2)
$$
Zero order in $a_{\mu j}$ term
$$
\bar{\Phi}_0=\Delta_0^2(\tau+2\kappa)+\frac{1}{2}\beta_{12345}\Delta_0^4                                \eqno(3)
$$
represents a gain of energy of the polar phase with respect to the normal and determines the temperature dependence of the amplitude $\Delta_0$: $\Delta_0^2=-\frac{\tau+2\kappa}{\beta_{12345}}$. Explicit form of the second order term depends on a choice of  gauge of $A_{\mu j}^0$. It is convenient to take $\varphi=0$, so that $A_{\mu j}^0$ is a real matrix.  The expected transition is due to the occurrence of two projections of angular momentum $l_z=\pm 1$ previously suppressed by the anisotropy. It means that only components of $a_{\mu j}$  transverse to $m_j$ are essential, and the condition  $a_{\mu j}m_j=0$ has to be imposed. With this simplification:
$$
\bar{\Phi}_2=(\tau-\kappa)a_{\mu j}a_{\mu j}^*+\frac{1}{2}\Delta_0^2\{\beta_1(a_{\mu j}a_{\mu j}+a_{\mu j}^*a_{\mu j}^*)+
2\beta_2a_{\mu j}a_{\mu j}^*+
$$
$$
\beta_3d_{\mu}d_{\nu}(a_{\mu j}a_{\nu j}+a_{\mu j}^*a_{\nu j}^*)+2\beta_{45}d_{\mu}d_{\nu}a_{\mu j}a_{\nu j}^*\}                                           \eqno(4)
$$
All experimentally observed transitions take place at $|\tau|\sim.1$, where Eq.(1) is still a good approximation for $\bar{\Phi}$. For this reason corrections of the order of $\tau$, originating from the higher order terms in the expansion of $\bar{\Phi}$ over $A_{\mu j}$  are neglected and the fourth order in $a_{\mu j}$ terms have the same form as the analogous terms in Eq.(1) with the substitution of  $a_{\mu j}$ instead of $A_{\mu j}$:
$$
\bar{\Phi}_4=\frac{1}{2}\sum_s\beta_sI_s(a_{\mu j},a_{\nu l}^*).        \eqno(5)
$$

The polar phase preserves its stability (or meta-stability) until $\bar{\Phi}_2$  is positively definite with respect to  $a_{\mu j}$. Coefficients in $\bar{\Phi}_2$ are determined by the more symmetric (polar) phase. In particular, $d_{\mu}$ and $m_j$ are symmetry axes in spin and orbital spaces respectively, so that essentially different are components of $a_{\mu j}$ parallel and perpendicular to  $d_{\mu}$. Of possible orientations in the orbital space only perpendicular to $m_j$ components are essential, as it was explained before. Because of the fixed gauge of the polar phase $\bar{\Phi}_2$ is not gauge invariant with respect to $a_{\mu j}$. On the other hand $\bar{\Phi}_2$ is $T$-invariant and transition to $T$-even $2b_{\mu j}=a_{\mu j}+a_{\mu j}^*$ and  $T$-odd $2ic_{\mu j}=a_{\mu j}-a_{\mu j}^*$ combinations of $a_{\mu j}$ diagonalizes $\bar{\Phi}_2$:
$$
\bar{\Phi}_2=\Lambda_1(\tau,\kappa)(\delta_{\mu\nu}-d_{\mu}d_{\nu})b_{\mu j}b_{\nu j}+[\Lambda_1(\tau,\kappa)+(\beta_{45}-2\beta_1-\beta_3)\Delta_0^2]d_{\mu}d_{\nu}c_{\mu j}c_{\nu j}+
$$
$$
\Lambda_2(\tau,\kappa)d_{\mu}d_{\nu}b_{\mu j}b_{\nu j}+[\Lambda_2(\tau,\kappa)-(\beta_{45}+2\beta_1+\beta_3)\Delta_0^2](\delta_{\mu\nu}-d_{\mu}d_{\nu})c_{\mu j}c_{\nu j}.   \eqno(6)
$$
Here $\Lambda_1=\tau-\kappa+\beta_{12}\Delta_0^2$ and $\Lambda_2=\tau-\kappa+\beta_{12345}\Delta_0^2$. So, for each $j$ there are four different variables: real and imaginary parts of parallel and orthogonal to $d_{\mu}$ components of $a_{\mu j}$. In principle there may be four different continuous transitions from the polar phase into a less symmetric one. Each of the transitions takes place  when coefficient in front of the corresponding second order term changes sign, e.g. the perpendicular to $d_{\mu}$ component of the real part of  $a_{\mu j}$ can occur at $\tau$ determined by the condition $\Lambda_1(\tau,\kappa)=0$. Mostly important is the transition with the highest $\tau$ of four.

When coefficients $\beta_s$ have their weak coupling values the combinations $\beta_{45}\equiv\varepsilon=0$ and $2\beta_1+\beta_3\equiv\nu=0$. In this limit instead of four different transition temperatures there are two, they are determined by the conditions: $\Lambda_1(\tau_1,\kappa)$=0 and $\Lambda_2(\tau_2,\kappa)$=0, both are doubly degenerate.
The degeneracy is a manifestation of the "hidden symmetry", which exists in the weak coupling limit for the Equal Spin Pairing (ESP) states. For such states the quantization axis of spin can be chosen so that the condensate of Cooper pairs contains only pairs with spin projections $\pm 1$. The BCS Hamiltonian in that case does not couple condensates with different spin projections and these two condensates can be treated as independent. In particular they can have different orientations of orbital parts of the order parameter and differen complex phases. Strong coupling corrections install a coupling between the two condensates and lift the corresponding degeneracy. The "hidden symmetry" was previously discussed in a context of classification of  collective modes in the ABM-phase Refs.\cite{VH,VW}.

Condition $\Lambda_2(\tau,\kappa)$=0 leads to $\kappa=0$. At a finite $\kappa$ there is no transition, resulting in occurrence of $b_{\mu j}$ parallel to $d_{\mu j}$ and $c_{\mu j}$ perpendicular to $d_{\mu j}$. Another condition  $\Lambda_1(\tau,\kappa)$=0 has solution $\tau\equiv\tau_B=\kappa(1+3\beta_{12}/\beta_{345})$. In the weak coupling limit below this $\tau$ the incremental order parameter is a linear combination of $b_{\mu j}$ perpendicular to $d_{\mu j}$ and $c_{\mu j}$  parallel to $d_{\mu j}$. When the strong coupling corrections are restored, they lift this degeneracy so that $\tau_B$ relates only to $b_{\mu j}$ perpendicular, while the parallel components $c_{\mu j}$ can occur below  $\tau_A=\kappa(3\beta_{245}-\beta_{13})/2\beta_{13}$.
 Difference between the two transition temperatures can be expressed in terms of parameters $\varepsilon,\nu$, introduced above:
$\tau_A-\tau_B=\frac{3\kappa\beta_{12345}}{2\beta_{13}\beta_{345}}(\varepsilon-\nu)$. Depending on a sign of the difference  $(\varepsilon-\nu)$ one or another type of the order parameter is favored. The strong coupling corrections ``transfer'' the orbital anisotropy in the spin space.  If $d_{\mu}$ is taken as a quantization axis, components, parallel to  $d_{\mu}$ correspond to $s_z=0$ and perpendicular -- to $s_z=\pm 1$.   At  $\varepsilon>\nu$ the favored incremental order parameter is a combination of projections $s_z=\pm 1$ so that the full order parameter is:
$$
A^B_{\mu j}=\Delta_0 d_{\mu}m_j+\Delta_2e_{\mu}l_j+\Delta_3f_{\mu}n_j,                                    \eqno(7)
$$
where $n_j$ and $l_j$ are two mutually orthogonal vectors, forming together with $m_j$  basis in the orbital space, $\Delta_2$ and $\Delta_3$ are real amplitudes. Minimization of $\bar{\Phi}$ with respect to the amplitudes $\Delta_0,\Delta_2,\Delta_3$ renders:
$\Delta_2^2=\Delta_3^2=-\frac{\tau-\tau_B}{3\beta_{12}+\beta_{345}}$, $\Delta_0^2=-\frac{3\kappa}{\beta_{345}}+\Delta_2^2$ with the energy gain $\Phi_B-\Phi_p=-\frac{\beta_{345}}{\beta_{12345}(3\beta_{12}+\beta_{345})}(\tau-\tau_B)^2$.
This is the order parameter of the distorted BW-phase, it is specified by two amplitudes  $\Delta_0$ and $\Delta_2$ with different temperature dependences \cite{AI,FS}.

For conditions of the experiments \cite{askh} on cooling from the polar phase the distorted ABM-phase occurs first. It means that  the opposite inequality is met $\varepsilon<\nu$. In this case the increment $c_{\mu j}$  is  parallel to  $d_{\mu}$, it corresponds to $s_z=0$ and the resulting order parameter is:
$$
A^A_{\mu j}=\Delta_0 d_{\mu}m_j+i\Delta_1d_{\mu}n_j.                                    \eqno(8)
$$
Here, $\Delta_1$ is a real amplitude.  The  temperature dependences of $\Delta_0,\Delta_1$ are found by minimization of  $\bar{\Phi}$:
$\Delta_1^2=-\frac{\tau-\tau_A}{2\beta_{245}}$, $\Delta_0^2=-\frac{3\kappa}{2\beta_{13}}-\frac{\tau-\tau_A}{2\beta_{245}}$.
In comparison with the polar phase the new phase has lower thermodynamic potential. The gain is
$\Phi_A-\Phi_p=-\frac{\beta_{13}}{2\beta_{245}\beta_{12345}}(\tau-\tau_A)^2$  \cite{FS}.
The distorted BW-phase Eq. (8) in conditions of the experiments \cite{askh} is reached via the first order transition. The temperature $\tau_B$ preserves its meaning of the upper limiting temperature for existence of this phase and it determines temperature dependencies of the amplitudes $\Delta_2^2$ and $\Delta_3^2$.
\section{Axi-planar phase}
In the distorted ABM-phase $d_{\mu}$ is still a symmetry axis in spin space. Further lowering of this symmetry via a continuous phase transition is possible at cooling when the suppressed perpendicular projection  $b_{\mu j}$ comes to effect. Analysis of stability of $A^A_{\mu j}$ with respect to $b_{\mu j}$ along the lines of a previous section with the order parameter of a form $A_{\mu j}=A^A_{\mu j}+b_{\mu j}$ renders instead of Eq. (6):
$$
\bar{\Phi}_2=[\tau-\kappa+\beta_{12}\Delta_0^2+(\beta_{234}-\beta_{15})\Delta_1^2](b_{\mu j}n_j)(b_{\mu i}n_i)+
$$
$$
[\tau-\kappa+\beta_{12}\Delta_0^2+(\beta_1-\beta_2)\Delta_1^2](b_{\mu j}l_j)(b_{\mu i}l_i).                \eqno(9)
$$
Combination in front of $(b_{\mu j}n_j)(b_{\mu i}n_i)$ is positive, but that in front of $(b_{\mu j}l_j)(b_{\mu i}l_i)$ changes sign at $\tau=\tau_{AP}$, where
$$
\tau_{AP}=-\frac{\kappa}{2}+\frac{3\kappa}{2}\frac{\beta_{245}}{\beta_{13}}\frac{\nu}{\varepsilon}         \eqno(10)
$$
It is a temperature of continuous transition in a phase with finite  $(b_{\mu j}l_j)(b_{\mu i}l_i)$. Its dependence on parameters   $\nu$ and $\varepsilon$ in a limit $\nu\rightarrow 0$ and $\varepsilon\rightarrow 0$ is singular.  From
$\tau_{AP}-\tau_A=\frac{3\kappa}{2}\frac{\beta_{245}}{\beta_{13}}(\frac{\nu}{\varepsilon}-1)$ follows that at $\nu>\varepsilon$ $\tau_{AP}<\tau_A$ since. Transition temperature $\tau_{AP}$ is not far from $\tau_A$ if  $\varepsilon\approx \nu$
At $\varepsilon\approx 3\kappa\nu$  $\tau_{AP}$ moves to low temperatures, well beyond the limit of applicability of expansion (1). Below $\tau_{AP}$ the order parameter is that of the axi-planar phase:
$$
A_{\mu j}=\Delta_0d_{\mu}m_j+i\Delta_1d_{\mu}n_j+\Delta_2e_{\mu}l_j.  \eqno(11)
$$
Minimization of thermodynamic potential over  $\Delta_0,\Delta_1,\Delta_2$ renders following equations for the amplitudes:
$$
[\beta_{12345}\Delta_0^2+(\beta_{245}-\beta_{13})\Delta_1^2+\beta_{12}\Delta_2^2+(\tau+2\kappa)]\Delta_0=0,  \eqno(12)
$$
$$
[(\beta_{245}-\beta_{13})\Delta_0^2+\beta_{1-5}\Delta_1^2+(\beta_2-\beta_1)\Delta_2^2+(\tau-\kappa)]\Delta_1=0,     \eqno(13)
$$
$$
[\beta_{12}\Delta_0^2+(\beta_2-\beta_1)\Delta_1^2+\beta_{12345}\Delta_2^2+(\tau-\kappa)]\Delta_2=0,    \eqno(14)
$$
In case of $\nu>\varepsilon$ solutions of these equations reproduce the sequence of phase transitions at cooling from $\tau=-2\kappa$. At $\tau_A<\tau<-2\kappa$ a stable solution is $\Delta_1=0$, $\Delta_2=0$ and from Eq. (12)
$\Delta_0^2=-\frac{\tau+2\kappa}{\beta_{12345}}$. At $\tau=\tau_A$ $\Delta_1$ starts to grow, indicating second order phase transition in the distorted ABM phase, as discussed at the end of the previous section. If  $\tau_{AP}$ is within the limits of applicability of Ginzburg and Landau expansion the distorted ABM phase remains stable in the interval $\tau_{AP}<\tau<\tau_A$.
Below $\tau_{AP}$ $\Delta_2$ becomes finite and Eqns. (12)-(14) render solution, corresponding to the axi-planar phase:
$$
\Delta_2^2=\frac{\varepsilon\beta_{13}(\tau_{AP}-\tau)}{(\nu+\varepsilon)\beta_2\beta_3+\varepsilon\nu\beta_{23}+
\varepsilon^2\beta_{13}},                                       \eqno(15)
$$
$$
\Delta_1^2=-\frac{3\kappa}{4\beta_{13}}\frac{\nu-\varepsilon}{\varepsilon}+\frac{(\beta_3+
\varepsilon)\nu}{2\beta_{13}\varepsilon}\Delta_2^2,               \eqno(16)
$$
$$
\Delta_0^2=\frac{\nu+\varepsilon}{2\beta_{13}\varepsilon}\left(-\frac{3\kappa}{2}+\beta_3\Delta_2^2\right).       \eqno(17)
$$
It follows from Eq. (15) that the axi-planar phase can exist only if $\varepsilon>0$.  Experimentally it can be detected by CW NMR method. As it was discussed in Ref. \cite{FS} when magnetic field is perpendicular to the anisotropy axis transverse  NMR shift is zero for the distorted ABM-phase, but it is finite and proportional to $\Delta_2^2$ in the axi-planar phase. The transition could be detected also by a jump of specific heat at cooling of the distorted ABM phase.

Both  $\Delta_2^2$ and $\tau_{AP}$ are very sensitive to values of parameters $\nu$ and $\varepsilon$, which for superfluid $^3$He in nematically ordered aerogel are poorly known. For the moment it is difficult to make even estimations of their values. In the published NMR data \cite{askh} the distorted ABM phase remains meta-stable till $\approx .7T_c$, when it jumps to the low temperature phase, which is identified as the distorted BW phase. There is no indication of continuous transition of the distorted ABM in the axi-planar phase. More detailed theoretical discussion of properties of the axi-planar phase could be appropriate if such indications would be available.
\section{Discussion}
Nematically ordered aerogel turned out to be an efficient tool for ``decomposition'' of the order parameter of  superfluid $^3$He in its constituents. Phenomenological analysis shows that in principle there are more possible phase diagrams than it is observed in real $^3$He and found in the model calculation \cite{AI}. The first superfluid phase realized immediately below  T$_c$ is always the polar phase. Its symmetry is higher, than that of the most stable at low temperature phases ABM and BW. It means that these phases can be reached in steps, via one or more phase transitions. The concrete route depends on particular values of phenomenological parameters $\beta$. Of importance are two combinations of these parameters $\varepsilon=\beta_4+\beta_5$ and $\nu=2\beta_1+\beta_3$. If $\varepsilon>\nu$ the distorted BW phase can form at cooling via second order transition directly from the polar phase. Such scenario is admitted by symmetry, but not realized either in the experiment \cite{askh} or in the microscopic calculations \cite{AI}. Situation $\varepsilon<\nu$ corresponds to the observed sequence of the phase transitions on cooling - the distorted ABM phase forms from the polar phase via continuous transition. If $\varepsilon>0$ and  $\varepsilon\approx\nu$ further continuous transition - in the axi-planar phase is possible. This transition is also admitted by symmetry, but not observed. Unfortunately there are no efficient tools for tuning parameters $\beta$ so that not all possible phase diagrams can be realized in real $^3$He, but phenomenological description can be used as a framework for a systematic description of experimental data for the realized scenario in a vicinity of transition of $^3$He in the superfluid state.

I thank V. V. Dmitriev and E.V.Surovtsev for stimulating discussions and useful comments. This research was supported in part by the Russian Foundation for Basic Research under projects  11-02-00357-a and 14-02-00054-a.

\end{document}